\documentclass[pra,aps,showpacs,epsf,twocolumn]{revtex4}

\usepackage{graphicx}
\usepackage{epsfig}
\usepackage{color}

\begin{document}

\title{Degenerate atom-molecule mixture in a cold Fermi gas}
\author{S.J.J.M.F. Kokkelmans$^1$, G.V. Shlyapnikov$^{1,2,3}$, and C. Salomon$^1$}
\affiliation{{$^1$Laboratoire Kastler Brossel, Ecole
Normale Sup\'erieure, 24 rue Lhomond, 75231 Paris 05,
France}\\{$^2$FOM Institute AMOLF, Kruislaan 407, 1098 SJ
Amsterdam, The Netherlands}\\{$^3$Russian Research Center
Kurchatov Institute, Kurchatov Square, 123182 Moscow, Russia}}
\date{\today}

\begin{abstract}
We show that the atom-molecule mixture formed in a degenerate atomic Fermi gas with interspecies repulsion near a Feshbach resonance, constitutes a peculiar system where the atomic component is almost non-degenerate but quantum degeneracy of molecules is important. We develop a thermodynamic approach for studying this mixture, explain experimental observations and predict optimal conditions for achieving molecular BEC.
\end{abstract}


\maketitle

Interactions between particles play a crucial role in the behavior of
degenerate quantum gases. For instance, the sign of the effective mean field
interaction determines the stability of a large Bose-Einstein condensate
(BEC), and the shape of such a condensate in a trap can be significantly
altered from its ideal gas form~\cite{dalfovo}. In degenerate Fermi gases the
effects of mean field interactions are usually less pronounced in the size and
shape of the trapped cloud, and these quantities are mostly determined by
Fermi statistics. The strength of the interactions, however,
can be strongly increased by making use of a Feshbach
resonance~\cite{feshbach,tiesinga}, and then the situation changes.

Recent experiments present two types of measurement of the interaction
energy in a degenerate two-component Fermi gas near a Feshbach
resonance~\cite{gehm,regal,bourdel,gupta}.
At JILA~\cite{regal} and at MIT~\cite{gupta} the mean field energy was found
from the frequency shift of an RF transition for one of the atomic states.
The results are consistent with the magnetic field dependence of
the scattering length $a$, the energy being positive for $a>0$ and negative for
$a<0$. In the Duke \cite{gehm} and ENS \cite{bourdel} experiments with $^6$Li,
the results are quite different. The interaction energy was obtained from the
measurement of the size of an expanding cloud released from the trap.
A constant ratio of the interaction to Fermi energy, $E_{\rm
int}/E_F\approx -0.3$, was found around resonance, irrespective
of the sign of $a$~\cite{gehm,bourdel}. It was explained in Ref.~\cite{gehm}
by claiming a universal behavior in this strongly
interacting regime~\cite{heiselberg}.  The ENS studies in a wide range of magnetic
fields~\cite{bourdel} found that
$E_{\rm int}$ changes to a large positive value when $a$ is
tuned positive, but only at a field strongly shifted from resonance.

In contrast to the JILA~\cite{regal} and MIT~\cite{gupta} studies providing a
direct measurement of the mean field interaction energy, the Duke~\cite{gehm}
and ENS~\cite{bourdel} experiments measure the influence of the interactions on
the gas pressure. An interpretation of the ENS experiment involves the
creation of weakly bound molecules via three-body recombination at a positive
$a$~\cite{bourdel}. Far from resonance, the binding energy of the produced molecules
and, hence, their kinetic energy are larger than the trap depth and the molecules
escape from the trap. The interaction energy is then determined by the repulsive
interaction between atoms and is positive~\cite{bourdel}.
Close to resonance, the three-body recombination is efficient \cite{petrov} and
the molecules remain trapped as their binding energy $\epsilon_B$
is smaller than the trap depth~\cite{bourdel,petrov}.
They come to equilibrium with the atoms, reducing the pressure
in the system.

Away from resonances, the interaction strength is proportional to $a$, and is given by $g=4\pi \hbar^2 a/M$, with $M$ the atom mass. Close to resonance this relation is not valid, as the value of $|a|$ diverges to
infinity and the scattering process strongly depends on the collision energy.
For Boltzmann gases, already in the 1930's, Beth and Uhlenbeck~\cite{beth}
calculated the second virial coefficient by including both the scattering and bound states for the relative motion of pairs of atoms~\cite{LLVol5}.
A small interaction-induced change of the pressure in this approach
is negative on both sides of the resonance~\cite{privcomm,ho}.

However, current experiments are not in the Boltzmann regime.
In this letter we show that the atom-molecule mixture formed in a cold
atomic Fermi gas, constitutes a peculiar
system in which the atomic component is almost non-degenerate, whereas
quantum degeneracy of the molecules can be very important. This
behavior originates from a decrease of the atomic fraction with temperature.
It is present even if the initial Fermi gas is strongly degenerate in which case
almost all atoms are converted into molecules. We develop a thermodynamic
approach for studying this mixture, predict optimal conditions
for achieving molecular BEC, and properly describe the interaction
effects as observed at ENS~\cite{bourdel}.

We assume that fermionic atoms are in equilibrium with weakly bound (bosonic) molecules formed in the recombination process. The molecules are treated as point bosons. Atom-molecule and molecule-molecule interactions are omitted at first, and will be discussed in a later stage. For a large scattering length $a>0$,
the binding energy of the weakly bound molecules is
$\epsilon_B=\hbar^2/(Ma^2)$, and their size is
roughly given by $a/2$. For treating them as point bosons,
this size should be smaller than the mean
interparticle separation. This requires the inequality
$n(a/2)^3<1$, which at
densities $n\!\approx\!10^{13}$ cm$^{-3}$ is satisfied
for $a<18000 a_0$, and excludes a narrow vicinity of the Feshbach resonance.

The presence of molecules reduces the number
of particles in the atomic component and to an essential extent lifts its
quantum degeneracy. The molecular chemical potential is negative in the absence of
atom-molecule and molecule-molecule mean field, and
thermal equilibrium between atoms and molecules requires a negative
chemical potential for the atoms.
We thus assume {\it a priori} that the occupation
numbers of the states of atoms are small.
This proves to be the case at any temperature,
except for very low $T$ where the atomic fraction is negligible.
Under these conditions we omit pairing correlations between the atoms, which
are important for describing a crossover from the BCS to BEC
regime~\cite{nozieres,randeria,ohashi,milstein} and can be
expected even in the non-superfluid state.

\begin{figure}[t]
\begin{center}
\epsfig{file=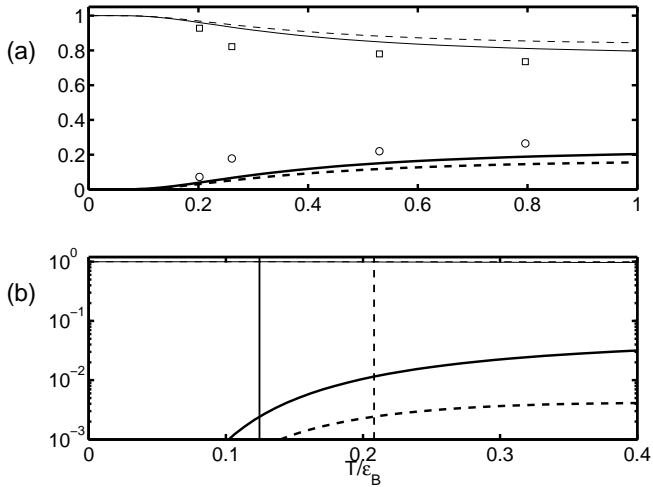, width=\linewidth} \caption{
\label{fig1}
Fraction of unbound atoms $n_a/n$ (lower curves, bold) and fraction of atoms bound
into molecules, $2n_m/n$, (upper curves) versus $T/\epsilon_B$:
a) $n\Lambda_T^3=2.5$, squares and circles show the ENS data~\cite{cubizolles};
b) $n\Lambda_T^3=14.8$, and the vertical lines indicate the onset of
molecular BEC. Dashed curves are obtained including atom-molecule and
molecule-molecule interactions.} \end{center}
\end{figure}

Assuming equal densities of the atomic components, labeled as $\uparrow$ and $\downarrow$, their chemical potentials are $\mu_{\uparrow}=\mu_{\downarrow}=\mu$,
where $\mu$ is the chemical potential of the system as a whole.
The molecular chemical potential is
$\mu_m=-\epsilon_B+\tilde\mu_m$, with $\tilde\mu_m\leq 0$ being the chemical
potential of an ideal gas of bosons with the mass $2M$. The condition of
thermal equilibrium, $\mu_{\uparrow}+\mu_{\downarrow}=\mu_m$, then reads
\begin{equation}      \label{mu}
2\mu=-\epsilon_B+\tilde\mu_m.
\end{equation}
From Eq.(\ref{mu}) we will obtain the number of molecules $N_m$
and the number of atoms $N_a$ for given temperature $T$ and total
number of atomic particles $N=N_a+2N_m$.
This requires us to obtain the expression for the occupation
numbers of the atoms and the dependence of $\mu$ on $N_a$.

The main difficulty with constructing a thermodynamic approach for
the degenerate molecule-atom mixture is related to the resonance
momentum-dependent character of the atom-atom interactions. This difficulty is circumvented for small occupation numbers of the
atoms. Then, even at resonance, the interaction energy
is equal to the mean value of the interaction potential for a given relative momentum of a colliding pair, averaged over the momentum distribution. In this respect, the interaction problem becomes similar
to the calculation of the total energy of a heavy impurity as caused
by its interactions with the surrounding electrons in a metal~\cite{mahan}.
This approach leads to a relation between the collision-induced shift of
the energy levels of particles in a large spherical box, and the
scattering phase shift. Adding the integration over the states of the
center of mass motion for pairs of atoms, we find that the total
energy of interatomic interaction is equal to $\sum_{{\bf k}
\bf{k}'} g_{{\bf k}{\bf k}'}\nu_{\uparrow}(k,\mu,T)
\nu_{\downarrow}(k',\mu,T)/V$, where
$\nu_{\uparrow}$ and $\nu_{\downarrow}$ are occupation numbers of
single-particle momentum states, and $V$ is the volume (cf.~\cite{mahan}).
The momentum-dependent coupling constant is given by
\begin{equation}     \label{gk}
g_{{\bf k}{\bf k}'}=-\frac{4\pi\hbar^2}{M}\frac{\delta(|{\bf k}-{\bf k}'|/2)}
{|{\bf k}-{\bf k}'|/2}.
\end{equation}
The phase shift $\delta$ is expressed through the relative momentum
$q=|{\bf k}-{\bf k}'|/2$ and the
scattering length $a$ as $\delta=-\arctan{qa}$. In the limit of $q|a|\ll 1$,
Eq.(\ref{gk}) transforms into the ordinary coupling constant
$g=4\pi\hbar^2a/M$.

As we have $\nu_{\uparrow}(k,\mu,T)=\nu_{\downarrow}(k,\mu,T)\equiv\nu_k$,
the total energy of the atomic component and the number of particles in this
component can be written in the form
\begin{eqnarray}    \label{Ea}
E_a=\sum_{{\bf k}}\frac{\hbar^2k^2}{M}\nu_k+\sum_{{\bf k}{\bf
k}'}\frac{g_{{\bf k}{\bf k}'}}{V}\nu_k\nu_{k'};\,\,\,\,
N_a=2\sum_{{\bf k}}\nu_k.
\end{eqnarray}
In our mean-field approach, the entropy of the atoms is given by
the usual combinatorial expression~\cite{LL9}:
\begin{equation}      \label{Sa}
S_a=-2\sum_{{\bf k}}[\nu_k\ln{\nu_k}+(1-\nu_k)\ln{(1-\nu_k)}].
\end{equation}
Equations (\ref{Ea}) and (\ref{Sa}) immediately lead to an expression for the
atomic grand potential $\Omega_a=E_a-TS_a-\mu N_a$. Then, using the relation
$N_a=-(\partial \Omega_a/\partial \mu)_{T,V}$,
we obtain for the occupation numbers of atoms:
\begin{equation}     \label{nu}
\nu_k=[\exp\{(\epsilon_k-\mu)/T\}+1]^{-1},
\end{equation}
where $\epsilon_k=\hbar^2k^2/2M+U_k$, and $U_k=\sum_{\bf k'} g_{{\bf
k}{\bf k}'}\nu_{k'}/V$ is the mean field acting on the atom with
momentum ${\bf k}$. Accordingly, the expression for the grand potential and
pressure of the atomic component reads:
\begin{equation}     \label{Pa}
\Omega_a=-P_aV=\sum_{{\bf k}}[2T\ln{(1-\nu_k)}-U_k\nu_k].
\end{equation}

This set of equations is completed by the relation between the density
of  bosonic molecules and their chemical potential. In the absence of molecular
BEC we have:
\begin{equation}    \label{Nm}
n_m=(\sqrt{2}/\Lambda_T)^{3/2}g_{3/2}(\exp{(\tilde\mu_m/T)},
\end{equation}
where $g_{\alpha}(x)=\sum_{j=1}^{\infty}x^j/j^{\alpha}$, and
$\Lambda_T=(2\pi\hbar^2/MT)^{1/2}$ is the thermal de Broglie
wavelength for the atoms. For $n_m\Lambda_T^3>7.38$ the molecular fraction
becomes  Bose-condensed,  and we have $\tilde\mu=0$ and $\mu=-\epsilon_B/2$.
Similarly, the energy, entropy, and grand potential of the molecules are given
by usual equations for an ideal Bose gas~\cite{LLVol5}.

From Eqs.~(\ref{mu})-(\ref{Nm}) we obtain the fraction of unbound atoms
$n_a/n$ and the fraction of atoms bound into molecules, $2n_m/n$, as
universal functions of two parameters: $T/\epsilon_B$ and $n\Lambda_T^3$,
where $n$ is the total density of atomic particles.
The dependence of atomic and molecular fractions on $T/\epsilon_B$ for two
values of $n\Lambda_T^3$ is shown in Fig.\ref{fig1}. The molecular fraction
increases and the atomic fraction decreases with decreasing
$T/\epsilon_B$. Occupation numbers of the atoms are always small,
whereas quantum degeneracy of molecules is important. The dotted line in Fig.1b
indicates the onset of molecular BEC.

\begin{figure}[t]
\begin{center}
\epsfig{file=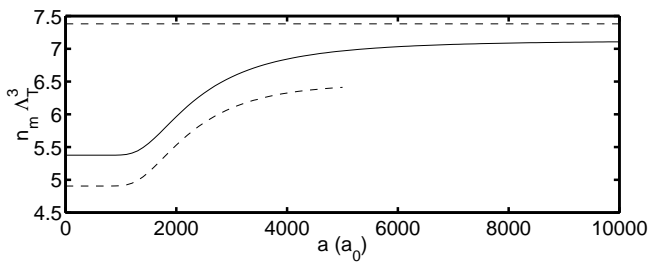, width=\linewidth}
\caption{  \label{fig2}
Molecular degeneracy parameter $n_m\Lambda_T^3$ under
adiabatic variation of $a$ for $^6$Li,
assuming $n\Lambda_T^3=15$ close to resonance. The dashed curve is
obtained including
atom-molecule and molecule-molecule interactions.
The horizontal dashed line shows the critical value for molecular BEC.
}
\end{center}
\end{figure}

This mixture was realized in the ENS experiment~\cite{bourdel}, where
the occupation numbers for the molecules were
up to $0.3$ and the molecular fraction was exceeding the atomic one.
In the recent studies~\cite{cubizolles,rudi,randy,regal1} almost all atoms were
converted into molecules by sweeping the magnetic field across the
resonance, and at ENS~\cite{cubizolles} the temperature was
within a factor of 2 from molecular BEC.
Remarkably, one can modify the molecular fraction
and degeneracy parameter $n_m\Lambda_T^3$ by
adiabatically tuning the atom-atom scattering length, as shown in Fig.2.
The decrease of $a$ increases the binding energy $\epsilon_B$
and the molecular fraction, and thus causes heating~\cite{cubizolles}. Close to resonance, $n_m\Lambda_T^3$ remains almost constant and then decreases due to heating.

\begin{figure}[b]
\begin{center}
\epsfig{file=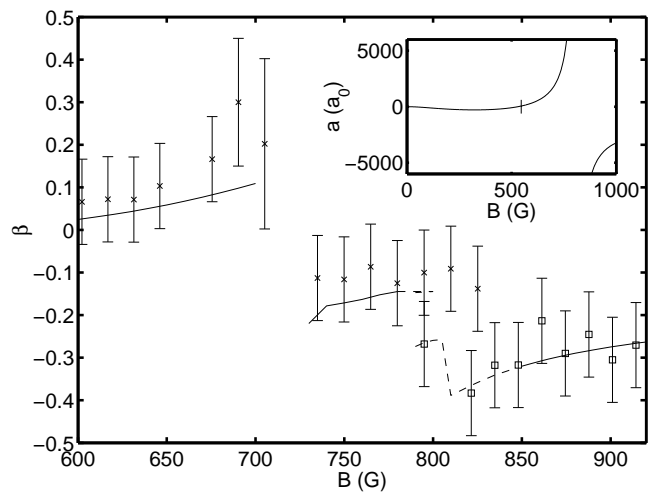, width=\linewidth}
\caption{  \label{fig3}
Calculated (solid line) and
measured~\protect\cite{bourdel} (squares and crosses) ratio $\beta$ of the
interaction to kinetic energy (see text). The calculated line for $B>790$G
is for experimental conditions $T=0.9E_F=3.4\mu$K and $n=3\times
10^{13}$cm$^{-3}$. For $B<700$G we take the averaged experimental
conditions $T=1.1E_F=2.4\mu$K and $n=1.3\times 10^{13}$cm$^{-3}$. For $700<B<800$G, we use the local conditions (see~\protect\cite{bourdel}). Inset: Scattering length as a function of
magnetic field.} \end{center} \end{figure}

The atom-molecule and molecule-molecule interactions are readily  included in our
approach for $a\ll\Lambda_T$, where the corresponding coupling constants
are $g_{am}=0.9g$ and $g_m=0.3g$~\cite{PS}.
In this limit the interactions provide an equal shift of the
chemical potential and single-particle energy $\epsilon_k$. For the atoms this
shift is $n_ag/2+n_mg_{am}$, where the first term is the atom-atom
contribution $U_k$. For the (non-condensed) molecules
the shift is $n_ag_{am}+2n_mg_m$. The entropy of the mixture is
given by the same expressions as in the absence of the interactions.
As seen in Fig.1 and Fig.2, the atom-molecule and molecule-molecule
interactions do not significantly modify our results. From Fig.2 one then
concludes that the conditions for achieving molecular BEC are optimal for
values of $a$ as low as possible while still staying at the plateau, as at
larger $a$ the interaction between the molecules can reduce the BEC transition
temperature~\cite{dalfovo}.

We now analyze the
interaction effect observed at ENS for trapped clouds in the
hydrodynamic regime~\cite{bourdel}. The experiment was done near
the Feshbach resonance located at the magnetic field $B_0=810$ G, and
the data results from two types of measurements of the size of the
cloud released from the optical trap. In the first
one, the magnetic field and, hence, the scattering length, are kept the same
as in the trap.
Therefore, the cloud expands with the speed of sound $c_s=\sqrt{(\partial
P/\partial\rho)_S}$, where $\rho=mn$ is the mass
density. The speed $c_s$ and, hence, the size of the expanding
cloud are influenced by the presence
of molecules and by the interparticle interactions.

In the second type of measurement, the magnetic field is first rapidly ramped
down and the scattering length becomes almost zero on a time scale
$t\sim 2\mu$s. This time
scale is short compared to the collisional time. Therefore, the
spatial distribution of the atoms remains the same as in the initial cloud,
although the mean field is no longer present.
At the same time, a rapid decrease of $a$ increases the binding energy
of molecules $\epsilon_B$. However, as the time $t\alt\hbar/\epsilon_B$,
they can not adiabatically follow to a deeper bound state and dissociate
into atoms which acquire kinetic energy. Thus the system expands symmetrically
as an ideal gas of $N$ atoms, with the initial density profile. The momentum distribution $f_k$ will be a sum of the
initial atomic momentum distribution and one that arises from the dissociated
molecules. The latter is found assuming an abrupt change of $a$ and, hence,
projecting the molecular wave function on a complete set of plane waves. This
gives rise to a distribution $c(q)$ for the relative momentum $q$. The
single-particle momentum distribution for the atoms produced out of
molecules results then from convoluting $c(|{\bf k}-{\bf k}'|/2)$ with the
molecular distribution function $\nu_m({\bf k}+{\bf k}')$ by integrating over
${\bf k}'$. One can establish a relation between the expansion velocity $v_0$
of this non-equilibrium system and the expansion velocity $c_0$ of
an ideal equilibrium two-component atomic Fermi gas which has the same density
and temperature: $4 \pi^3 n \int_0^{Mv_0/\hbar} dk k^2 f_k =
\int_0^{Mc_0/\hbar} dk k^2 \tilde\nu_k$, with $\tilde\nu_k$ being the
ideal-gas momentum distribution. Using the scaling approach \cite{kagan,str},
one can find that in the spherical case the velocity $c_0$ coincides with the
expansion velocity of the hydrodynamic Fermi gas in the absence of mean-field
interactions and, accordingly, is given by $c_0^2=5P_0/3\rho$, where
$P_0=2E_0/3V$  is the pressure.

The relative difference between the squared size of the expanding cloud in the
two described cases can be treated as the ratio of the interaction to kinetic
energy and called the interaction shift. This interaction shift is then given
by the relative difference between the two squared velocities:
$\beta=[c_s^2-v_0^2]/v_0^2$. Our results for this quantity are calculated for
experimental conditions and are presented in Fig.\ref{fig3}. The sound
velocity $c_s$ was obtained using the above developed approach including only
atom-atom interactions. The field region where  $n(a/2)^3>1$, is beyond the
validity of this approach and is shown by the dashed curve. In Fig.\ref{fig3}
we also show our previous results for fields $B>810$ G ($a<0$) and $B<700$ G
($0<a<2000 a_0$), where molecules are absent~\cite{bourdel}.

Our quantum-statistical approach gives a negative interaction shift on both sides
of the Feshbach resonance, in good quantitative agreement with the
experiment. Without molecules present, the interaction energy would jump to
positive values left from resonance, as can be seen from our calculation in
Ref.~\cite{bourdel}. This demonstrates that the apparent field shift from
resonance, where a sign-change in the interaction energy is observed, is an
indirect signature of the presence of molecules in the trap.

For high temperatures $T\gg E_F$ and small binding energy $\epsilon_B \ll T$,
we find that $\beta$ has a universal
behavior and is proportional to the second virial coefficient.
However, this only holds at high temperatures (cf.~\cite{ho}),
and at low $T$ the molecule-molecule interaction can strongly influence the
result. For $T$ approaching the temperature of molecular BEC, which is
$T_c\!\approx\!\hbar^2n^{2/3}/M\!\approx\!0.2E_F$, the atomic
fraction is already small and the sound velocity $c_s$ is determined by
the molecular cloud. For $a\ll\Lambda_T$ we find
$c_s^2=0.4T_c/M+ng_m/2M$, where the second term
is provided by the molecule-molecule interaction and is omitted in the high-$T$ approach.
The ratio of this term to the first one is $\sim 5(na^3)^{1/3}$. For $B=700$
G at densities of Ref.~\cite{bourdel}, it is equal to 1 and is expected to
grow when approaching the resonance.

Thus, except for a narrow region where $n |a|^3 \gg 1$, one can not speak of a
universal behavior of the shift $\beta$ on both sides of the resonance. The
situation depends on possibilities of creating an equilibrium atom-molecule
mixture. Moreover, at low temperatures the universality can be broken by the
molecule-molecule interactions.

We are grateful to T. Bourdel, J. Cubizolles, C. Lobo, and L. Carr for
stimulating discussions. This work was supported by the Dutch Foundations NWO
and FOM, by INTAS, and by the Russian Foundation for Fundamental Research.
S.K. acknowledges a Marie Curie grant MCFI-2002-00968 from the E.U.
Laboratoire Kastler Brossel is a Unit\'e de Recherche de l'Ecole Normale
Sup\'erieure et de l'Universit\'e Paris 6, associ\'ee au CNRS.

\end{document}